\documentclass[fleqn,12pt]{article}
\usepackage{epsfig}
\begin{document}
\textheight 22cm
\textwidth 15cm
\noindent
{\Large \bf Non-perturbative statistical theory of intermittency in ITG drift wave turbulence with zonal flows}
\newline
\newline
Johan Anderson\footnote{anderson.johan@gmail.com} and Eun-jin Kim
\newline
University of Sheffield
\newline
Department of Applied Mathematics
\newline
Hicks Building, Hounsfield Road 
\newline 
Sheffield
\newline
S3 7RH
\newline
UK
\newline
\newline
\begin{abstract}
\noindent
The probability distribution functions (PDFs) of momentum flux and zonal flow formation in ion-temperature-gradient (ITG) turbulence are investigated in two different models. The first is a general five field model ($n_i, \phi, T_i, T_e, v_{i \parallel}$) where a reductive perturbation method is used to derive dynamical equations for drift waves and a zonal flow. The second is a reduced two-field model ($\phi, T_i$) that has an exact non-linear solution (bipolar vortex soliton). In both models the exponential tails of the zonal flow PDFs are found with the same scaling ($PDF \sim \exp \{ - c_{ZF} \phi_{ZF}^3\}$), but with different coefficients $c_{ZF}$. The PDFs of momentum flux is however found to be qualitatively different with the scaling ($PDF \sim \exp \{ - c_{M} R^{s}\}$), where $s=2$ and $s=3/2$ in the five and two field models, respectively.
\end{abstract}
\newpage
\renewcommand{\thesection}{\Roman{section}}
\section{Introduction}
One of the main challenges in magnetic fusion research has been the prediction of the turbulent heat and particle transport originating from various micro-instabilities. The ion-temperature-gradient (ITG) mode is one of the main candidates for causing the anomalous heat transport in core plasmas of tokamaks~\cite{a1}. Significant heat transport can however be mediated by coherent structures such as streamers and blobs through the formation of avalanche like events of large amplitude, as indicated by recent numerical studies~\cite{a81}-~\cite{a83}. These events cause the deviation of the probability distribution functions (PDFs) from a Gaussian profile on which the traditional mean field theory (such as transport coefficients) is based. A crucial question in plasma confinement is thus the prediction of the PDFs of the transport due to these structures and of their formation~\cite{a3}-~\cite{a7}. Note that a renormalized perturbative theory can easily make an considerably large error in predicting PDFs where the coherent structures are crucial.

In previous papers the Hasegawa-Mima model of drift wave turbulence was used to show that the PDF tails of global momentum flux and heat flux are significantly enhanced over the Gaussian prediction~\cite{a3}-~\cite{a4}; this was later shown to hold also in ITG mode turbulence~\cite{a5}-~\cite{a7}. Specifically, the tails of PDF of global momentum flux and heat flux have been shown to be stretched exponential with the form $\sim \exp\{- c (R/R_0)^{3/2}\}$ ~\cite{a3}-~\cite{a7}, which is broader than a Gaussian. These results provided a novel explanation for exponential PDF tails of momentum flux found in recent experiments at CSDX at UCSD~\cite{a71}. Note that considerable transport can be mediated by rare events of high amplitude at the PDF tails even if the latter have a low amplitude. Since the PDF tails of momentum flux are enhanced over the Gaussian prediction in this case, these events are more likely to mediate considerable transport. In fact, it was found that for certain values of parameters, large events are crucial for transport. Specifically, the overall amplitude was shown to be larger in ITG than in HM turbulence for reversed modon speed ($U<0$). The PDF tail of zonal flow in ITG turbulence was also significantly increased compared to that in HM turbulence. Furthermore, zonal flows are shown to be generated more likely further from marginal stability, which will then regulate ITG turbulence, leading to a self-regulating system. Namely, while ITG turbulence is a state with high level of heat flux, it also generates stronger zonal flows that inhibit transport. This also suggested that stronger zonal flows are generated in ITG turbulence compared with ETG turbulence. It was also shown that shear flows can significantly reduce the PDF tails of Reynolds stress and zonal flow formation~\cite{a3}-~\cite{a7}. 

The purpose of the present paper is to investigate the likelihood of the formation of coherent structures by computing the PDF (tails) of zonal flow formation and the PDFs of momentum flux~\cite{a2} by using the two different ITG turbulence models. The first model is a widely applicable five field model ($n_i, \phi, T_i, T_e, v_{i \parallel}$) of the ITG turbulence coupled to an ion vorticity equation for the zonal flows. For the computation of the PDFs, a reductive perturbation method is used to obtain dynamical equations for drift waves and zonal flows. Due to the perturbation method, this model is only valid for weak drift wave and zonal flow electric potential. Moreover an exact non-linear solution in this model is not available that can be used as an ansatz for a coherent structure for the intermittent transport. However, the computation of the PDF tails requires only mean values over the coherent structures. The second is the two field ($\phi, T_i$) model which has an exact non-linear bipolar vortex soliton (modon) solution that will be utilized in the computing PDFs. Note that the two-field fluid model for the ITG mode~\cite{a8}-~\cite{a12} has been successful in reproducing both experimental~\cite{a51} and non-linear gyro-kinetic results~\cite{a52}. 

The theoretical technique used here is the so-called instanton method, a non-perturbative way of calculating the PDF tails. The PDF tail is first formally expressed in terms of a path integral by utilizing the Gaussian statistics of an external forcing with a short correlation time. An optimum path will then be associated with the creation of a coherent structure (among all possible paths) and the action is evaluated using the saddle-point method. In the second model this coherent structure is identified as the modon. The saddle-point solution of the dynamical variable $\phi(x,t)$ of the form  $\phi(x,t) = F(t)\psi (x)$ with $F(t)$ which is localized in time is called an instanton. Since the instanton exists during the formation of the coherent structure, the bursty event can be associated with the creation of a coherent structure. Note that the function $\psi(x)$ here represents the spatial form of the coherent structure.

One of our main results is that the PDF tail of the zonal flow formation  is $PDF \sim \exp \{ - c_{ZF j} \phi_{ZF}^3\}$, ($j = 1,2$ to denote the different constants in the two models), in agreement with earlier findings~\cite{a7}. Interestingly the predicted PDF tails of momentum flux are Gaussian when the feedback of zonal flow on the turbulence is incorporated. This result differs from what has theoretically been found earlier~\cite{a7} whereas it is in agreement with predictions from non-linear simulations of turbulence~\cite{a61}-~\cite{a62} and gyro-kinetic toroidal simulations~\cite{a91}-~\cite{a92}. The reason for this is that previously the zonal flow has been treated as passively evolving by the turbulence without the feedback loop to turbulence. This feedback gives a cubic non-linearity in Eq. (7) describing the ITG fluctuations which changes the characteristics of the turbulence. Physically, it is because the feedback of zonal flow regulates turbulence, limiting its growth from turbulence.

The result from both models support the conclusion that while ITG turbulence maintains high level of transport, this may be suppressed by shear flow. Zonal flows are also shown here to have an enhanced likelihood of the generation further from marginal stability which will then regulate the ITG turbulence (which is more prominent with increased shear flow) leading to a self-regulating system.

The paper is organized as follows. In Sec. II the drift wave - zonal flow system is presented and in Sec. III and IV the computation of the PDF and a numerical study of the five field model are presented. In Sec. V and VI the PDF tails are computed and a numerical study of the reduced two field model are outlined. The paper is concluded by Discussion and summary in Sec. VII.

\section{The drift wave - zonal flow system}
In the five-field model the ITG mode turbulence is modeled using the continuity and temperature equations for the ions and assuming the electrons to be Boltzmann distributed, closely following previous papers Ref.~\cite{a8}-~\cite{a12}. The ion continuity, temperature and parallel ion momentum equations then become,
\begin{eqnarray}
\frac{\partial n}{\partial t} - \left(\frac{\partial}{\partial t} - \alpha_i \frac{\partial}{\partial y}\right)\nabla^2_{\perp} \phi + \frac{\partial \phi}{\partial y} - \epsilon_n  g_i \frac{\partial}{\partial y} \left(\phi + \tau \left(n + T_i \right) \right)  \nonumber \\ + \nu \nabla^4 \phi =  - \left[\phi,n \right] + \left[\phi, \nabla^2_{\perp} \phi \right] + f_0\\
\frac{\partial }{\partial t}T_i - \frac{5}{3} \tau \epsilon_n g_i \frac{\partial T_i}{\partial y} + \left( \eta_i - \frac{2}{3}\right)\frac{\partial \phi}{\partial y} - \frac{2}{3} \frac{\partial n}{\partial t} =  \nonumber \\
- \left[\phi,T_i \right] + \frac{2}{3} \left[\phi,n \right] + S_{Ti}, \\
\frac{\partial v_{i \parallel}}{\partial t} = - \left( \frac{\partial \phi}{\partial z} + \tau \frac{\partial}{\partial z}(n + T_i)\right).
\end{eqnarray}
Eqs. (1)-(3) are closed by using the quasi-neutrality condition. Here $\left[ A ,B \right] = (\partial A/\partial x) (\partial B/\partial y) - (\partial A/\partial y) (\partial B/\partial x)$ is the Poisson bracket; $\nu = 0.78\times10^{-12}(n_0/T_{i0}^{3/2})(r/\bar{R})$ is a neoclassical damping; $f_0$ is a forcing; $n = (L_n/\rho_s) \delta n / n_0$, $\phi = (L_n/\rho_s) e \delta \phi /T_e$, $T_i = (L_n/\rho_s) \delta T_i / T_{i0}$ and $v_{i \parallel} = (L_n/\rho_s) \delta v_{i \parallel}/c_s$ are the normalized ion particle density, the electrostatic potential, the ion temperature, and the ion parallel velocity respectively. In equations (1) and (2), $\tau = T_i/T_e$, $\rho_s = c_s/\Omega_{ci}$ ($c_s=\sqrt{T_e/m_i}$, $\Omega_{ci} = eB/m_i c$). We also used $L_f = - \left( d ln f / dr\right)^{-1}$ ($f = \{n, T_i \}$), $\eta_i = L_n / L_{T_i}$, $\epsilon_n = 2 L_n / \bar{R}$ where $\bar{R}$ is the major radius and $\alpha_i = \tau \left( 1 + \eta_i\right)$. The perpendicular length scale and time are normalized by $\rho_s$ and $L_n/c_s$, respectively. The geometrical quantities are calculated in the strong ballooning limit ($\theta = 0 $, $g_i \left(\theta = 0 \right) = 1.0$) with $\omega_{\star} = k_y v_{\star} = \rho_s c_s k_y/L_n $. In the ion temperature equation [Eq. (2)] we have included a heat source $S_{Ti}$ given in Appendix. For the zonal flow Eq. (3) vanishes ($k_{\parallel} = 0$) and we cannot use Boltzmann distributed electrons. Thus we will employ the following ion vorticity equation [Eq. (4)] and simple electron energy equation [Eq. (5)]
\begin{eqnarray}
 \left( \frac{\partial }{\partial t} - \alpha_i \frac{\partial}{\partial y} - \nu \nabla_{\perp }^2 \right) \nabla_{\perp }^2 \phi + \epsilon_n g_i \left( (1+\tau ) \frac{\partial n}{\partial y} + \frac{\partial T_e}{\partial y}\right) = [\phi, \nabla_{\perp }^2 \phi ] \\
\frac{\partial T_e}{\partial t} = -\eta_e \frac{\partial \phi}{\partial y}
\end{eqnarray}
Here $\eta_e = L_n/L_{Te}$. The damping (with strength $\nu$) represents a neo-classical damping of the zonal flow. Note that the damping will affect the full dynamical system in Eq. (7)-(8). The simple electron energy equation is motivated by the fact that we are interested in the modes propagating in the ion drift direction.

In order to find the coupled system of equations for the ITG mode and the zonal flow we utilize the reductive perturbation method wherein the perturbed variable $\sigma$ is written 
\begin{eqnarray}
\sigma(x,y,t) & = & \sum_n \sum_{l \neq 0} \epsilon_0^{1+(2/3)n} \sigma_l^{[1+(2/3)n]}(x, \xi, \zeta) \times \exp\{ il(k_{\parallel}z + k_y y - \omega t)\} + c.c. \nonumber \\
& + & \sum_n \epsilon_0^{(4/3)+(2/3)n} \sigma_0^{[(4/3)+(2/3)n]}(x, \xi, \zeta) 
\end{eqnarray}
Here the coordinates are scaled as $\xi = \epsilon_0^{2/3}(y - \lambda t)$ and $\zeta = \epsilon^{4/3}t$ with a small expansion parameter $\epsilon_0 \sim \frac{e \phi}{T_e} \sim 10^{-2}$. The drift wave - zonal flow system then becomes
\begin{eqnarray}
C_1 \frac{\partial \tilde{\phi}_1}{\partial \zeta} + i C_2 \frac{\partial^2 \tilde{\phi}_1}{\partial \xi^2} + C_3 \tilde{\phi}_0 \tilde{\phi}_1 & = & -i \nu C_4 \tilde{\phi}_1 + f,\\
D_1 \frac{\partial \bar{\phi}_0}{\partial \zeta} + D_2 \frac{\partial \tilde{\phi}_0}{\partial \xi} & = & D_3 \frac{\partial |\tilde{\phi}_1|^2}{\partial \xi}. 
\end{eqnarray}
Here, the coefficients $C_1$, $C_2$, $C_3$, $C_4$, $D_1$, $D_2$ and $D_3$ are complex numbers whose forms are provided in the appendix; the variables with tilde and bar denote average over $x$. Note that if the time derivatives are neglected in Eqs. (7)-(8), a non-linear Schr\"{o}dinger (NLS) equation for the fluctuating potential is obtained~\cite{a12}.

\section{Non-perturbative calculation of zonal flow PDF in the five field model}
We calculate the PDF tails of momentum flux and zonal flow formation by using the instanton method. To this end, the PDF tail is expressed in terms of a path integral by utilizing the Gaussian statistics of the forcing $f$~\cite{a22}. The PDF of Reynolds stress and zonal flow formation can be defined as
\begin{eqnarray}
P(Z) & = &  \langle \delta(Z_0 - Z) \rangle \nonumber \\
& = & \int d \lambda  \exp(i \lambda Z) \langle \exp(-i \lambda Z_0) \rangle \nonumber \\
& = & \int d\lambda \exp(i \lambda Z) I_{\lambda},
\end{eqnarray}
where 
\begin{eqnarray}
I_{\lambda} = \langle \exp(-i \lambda Z_0) \rangle.
\end{eqnarray}
$I_{\lambda}$ in (9)-(10) can then be rewritten in the form of a path-integral as
\begin{eqnarray}
I_{\lambda} = \int \mathcal{D} \tilde{\phi}_1 \mathcal{D} \bar{\phi}_1 \mathcal{D} \tilde{\phi}_{0} \mathcal{D} \bar{\phi}_{0}  e^{-S_{\lambda}}.
\end{eqnarray}
In the following we assume that the zonal flow potential is stationary $\frac{\partial \bar{\phi}_0}{\partial t} = 0$ on the time-scale of the fluctuations. The effective action $S_{\lambda}$ in Eq. (11) can be then expressed as,
\begin{eqnarray}
S_{\lambda} & = & -i \int d \xi d\zeta \bar{\phi}_1 \left(C_1 \frac{\partial \tilde{\phi}_1}{\partial \zeta} + i C_2 \frac{\partial^2 \tilde{\phi}_1}{\partial \xi^2} + C_3 \tilde{\phi}_0 \tilde{\phi}_1 + i \nu C_4 \tilde{\phi}_1 \right) \nonumber \\
& + & \frac{1}{2} \int d\zeta d \xi d \xi^{\prime} \bar{\phi}_1(\zeta, \xi) \kappa(\xi-\xi^{\prime}) \bar{\phi}_1(\zeta, \xi^{\prime}) \nonumber \\
& + & i \lambda_2 \int d\zeta \tilde{\phi}_{0}(\zeta) \delta(\zeta) \nonumber \\
& - & i \int d\xi  d\zeta \bar{\phi}_{0}(\zeta)(D_2 \tilde{\phi}_0 - D_3 |\tilde{\phi}_1|^2 ).
\end{eqnarray}
Recall $\xi$ and $\zeta$ are the scaled spatial and time variables, respectively. To obtain Eq. (12) we have assumed the statistics of the forcing $f$ to be Gaussian with a short correlation time modeled by the delta function as
\begin{eqnarray}
\langle f(\xi, \zeta) f(\xi^{\prime}, \zeta^{\prime}) \rangle = \delta(\zeta-\zeta^{\prime})\kappa(\xi-\xi^{\prime}),
\end{eqnarray}
and $\langle f \rangle = 0$.
The delta correlation in time was chosen for the simplicity of the analysis. In the case of a finite correlation time the non-local integral equations in time are needed.

\section{The PDF tails and numerical studies for the five field model}
We have expressed the PDF in terms of a path-integral. An approximate value of this path-integral can be found for large values of the parameter $\lambda \rightarrow \infty$, by using a saddle point method. The action in Eq. (12) can be expressed using the instanton ansatz $\tilde{\phi}_1(\xi,\zeta) = \psi_1(\xi) F(\zeta)$ and $\tilde{\phi}_0(\xi,\zeta) = \psi_0(\xi) G(\zeta)$ as,
\begin{eqnarray}
S_{\lambda} & = & -i \int d\zeta \left(C_1 \dot{F}\bar{F}_1 + i C_2 k_{\xi}^2 F \bar{F}_1  + C_3 G F \bar{F}_2 + i \nu C_4 F \bar{F}_1 \right) \nonumber \\
& + & \frac{\kappa_0}{2} \int d\zeta \left( \bar{F}_1^2 + \bar{F}_2^2 \right) \nonumber \\
& + & i \lambda \Phi_0 \int d\zeta G(\zeta) \delta(\zeta) \nonumber \\
& - & i \int d\zeta \bar{F}_3 (D_2 G - R D_3 F^2 ).
\end{eqnarray}
Here $\kappa_0$ is the strength of the forcing $\kappa$. The parameter $k_{\xi} (\in \textbf{C})$ is the inverse length scale in the $\xi$ direction. In Eq. (14) we have defined the variables,
\begin{eqnarray}
\Phi_0 & = & \int d \xi \psi_0, \\
\bar{F}_1 & = & \int d \xi \bar{\phi}_1 \psi_1, \\
\bar{F}_2 & = & \int d \xi \bar{\phi}_1 \psi_1^2, \\
\bar{F}_3 & = & \int d \xi \bar{\phi}_0 \psi_0, \\
\bar{F}_4 & = & \int d \xi \bar{\phi}_0 \psi_1^2 = R \bar{F}_3.
\end{eqnarray}
Note that the relation between the conjugate variables $\bar{F}_3$ and $\bar{F}_4$ is based on the fact that the zonal flow is driven by the Reynolds stress and also that the parameter $R (\in \textbf{C})$ has to be determined in such a way to make Reynolds stress real. Since the saddle point action is determined by the extremum of the action, we require the first functional derivatives of the action to vanish;
\begin{eqnarray}
\frac{\delta S_{\lambda}}{\delta F} & = & -i ( -C_1 \dot{\bar{F}}_1 - i C_2 k_{\xi}^2 \bar{F}_1 + C_3 \bar{F}_2 G + i \nu C_4 \bar{F}_1) \nonumber \\
& - & 2 i R D_3 F \bar{F}_3 = 0 ,\\
\frac{\delta S_{\lambda}}{\delta \bar{F}_1} & = & -i(C_1 \dot{F} - i C_2 k_{\xi}^2 F + i \nu C_4 F ) + \kappa_0 \bar{F}_1 = 0,\\
\frac{\delta S_{\lambda}}{\delta \bar{F}_2} & = & -i C_3 F G + \kappa_0 \bar{F}_2 = 0,\\
\frac{\delta S_{\lambda}}{\delta \bar{F}_3} & = & D_2 G - R D_3 F^2 = 0,\\
\frac{\delta S_{\lambda}}{\delta G} & = & -i C_3 \bar{F}_2 F -i \bar{F}_3 D_2 + i \lambda \Phi_0 \delta(\zeta) = 0.
\end{eqnarray}
The initial conditions $F(0) = F_0$ and $G(0) = G_0$ are found from Eqs. (23)-(24) as,
\begin{eqnarray}
F_0 & = & \left( - \frac{i D_2 \Phi_0}{C_3^2 R D_3}\right)^{1/4} \lambda^{1/4},\\
G_0 & = & \frac{R D_3}{D_2}\left( - \frac{i D_2 \Phi_0}{C_3^2 R D_3}\right)^{1/2} \lambda^{1/2}.
\end{eqnarray}
The instanton solution is found from Eqs. (20) - (24) for $\zeta < 0$. Starting with Eq. (20) and then by substituting the conjugate variables in Eqs. (21) - (24) we obtain
\begin{eqnarray}
\ddot{F} = \frac{1}{2} \frac{d}{dF} \dot{F}^2 = \eta_1 F + 3 \eta_2 F^5,
\end{eqnarray}
where
\begin{eqnarray}
\eta_1 & = & \frac{1}{C_1^2}\left( -C_2^2 k_{\xi}^4 + 2 C_2 C_4 \nu k_{\xi}^2 - \nu^2 C_4^2 \right), \\
\eta_2 & = & \frac{1}{C_1^2} \left( \frac{C_3^2 R^2 D_3^2}{D_2^2}\right).
\end{eqnarray}
The integration of Eq. (27) in time, gives us the instanton solution as, 
\begin{eqnarray}
F(\zeta) & = & \sqrt{\frac{2 \eta_1 H}{1 - \eta_1 \eta_2 H^2}}, \\
H(\zeta) & = & H_0 \exp\{ 2 \sqrt{\eta_1} \zeta \}, \\
H_0 & = & \frac{F_0^2}{\eta_1 + \sqrt{\eta_1} \sqrt{\eta_2 F_0^4 + \eta_1}},
\end{eqnarray}
where $F_0$ is given in Eq. (25). The next task is to compute the action integrals and estimate the $\lambda$ dependence of $S_{\lambda}$ in the large $\lambda \rightarrow \infty$ limit. We start by substituting Eqs. (20) - (23) into the action in Eq. (14) keeping only the highest powers of $\lambda$,
\begin{eqnarray}
S_{\lambda} & \approx & i \int_{-\infty}^0 d \zeta \left( C_1^2 \dot{F}^2 + C_3^2 \frac{R^2 D_3^2}{D_2^2} F^6\right) \nonumber \\
& + & i \lambda \Phi_0 \frac{R D_3}{D_2} F_0^2 \\
& \approx &  i \int_{0}^{F_0} d F \left( 2 C_1^2 \dot{F} \right) + i \lambda \Phi_0 \frac{R D_3}{D_2} F_0^2 \nonumber \\
& \approx &  i \lambda^{3/2} \Phi_0 \frac{R D_3}{D_2} \left( - \frac{i D_2 \Phi_0}{C_3^2 R D_3}\right)^{1/2} \\
& = & i h_{ZF} \lambda^{3/2}.
\end{eqnarray}
Here,
\begin{eqnarray}
h_{ZF} = \Phi_0 \frac{R D_3}{D_2} \left( - \frac{i D_2 \Phi_0}{C_3^2 R D_3}\right)^{1/2}.
\end{eqnarray}
The PDF is then found from Eq. (9) by utilizing the saddle point method
\begin{eqnarray}
P(Z) & = & \int d \lambda e^{-i \lambda Z - S_{\lambda}} \\
& = & \int d \lambda  e^{-i \lambda Z - i h_{ZF} \lambda^{3/2}}.
\end{eqnarray}
We find a $\lambda_0$ such that $f(\lambda) = -i \lambda Z - i h_{ZF} \lambda^{3/2}$ attains its maximum and compute that value. This gives $\lambda_0 = \left( \frac{2Z}{3 h_{ZF}}\right)^2$ and 
\begin{eqnarray}
f(\lambda_0) = - \frac{4}{27 h_{ZF}^2} Z^3,
\end{eqnarray}
with the PDF
\begin{eqnarray}
P(Z) & \sim & e^{- \Theta Z^3}, \\
\Theta & = & \frac{4}{27 h_{ZF}^2}.
\end{eqnarray}
The PDF tails found in the five field model [Eq. (41)] has the same exponential form as in Ref.~\cite{a7}. Although no exact non-linear solution was used as an ansatz for the coherent structure the system, we still obtain the PDFs with the same exponential form. To elucidate on the quantitative differences the results between the two and five field models we show the PDF tails by using different values of temperature gradient ($\eta_i$) in Figures 1 and 3. 

In Figure 1 the PDF tails of zonal flow formation obtained in the five-field model [Eq.(40)] is presented for $\eta_i = 2.0 $ (red, dashed line), $\eta_i = 4.0 $ (blue, solid line) and $\eta_i = 6.0 $ (black, dashed-dotted line). Other parameter values are $\tau = 2.0$, $\epsilon_n = 1.0$, $g_i = 1$, $k_x = 0.3$, $k_y = 0.3$ and $\eta_e = 0.0$, $\kappa_0 = 3000$. It is interesting to notice that the PDF tails of zonal flow formation in Fig. 1 are suppressed in comparison with the results found in the reduced two field model shown in Figure 3A for similar parameter values.

\begin{figure}
  \includegraphics[height=.3\textheight]{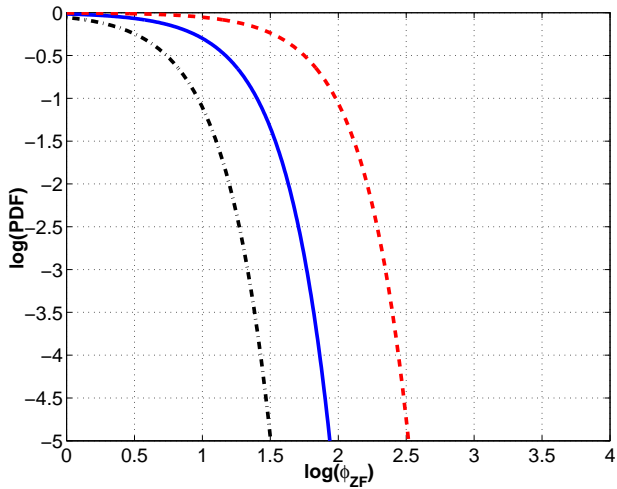}
  \caption{The PDF tail of zonal flow formation in the five field model for $\eta_i = 2.0$ (red, dashed line), $\eta_i = 4.0 $ (blue, solid line) and $\eta_i = 6.0 $ (black, dashed-dotted line). The parameters are $\tau = 2.0$, $\epsilon_n = 1.0$, $g_i = 1$, $k_x = 0.3$, $k_y = 0.3$, $\eta_e = 0.0$, $\kappa_0 = 3000$.}
\end{figure}

Note that the usefulness of the predicted PDF of zonal flow formation above is limited since the coefficient $\Theta$ in Eq. (40) is dependent on complex coefficients $C_1$, $C_2$, $C_2$, $C_3$, $C_4$, $D_1$, $D_2$ and $D_3$ from the original system of Eqs. (7) - (8). As is seen in the appendix due to the complexity of the coefficients it is rather difficult to give an explicit region where the analysis is useful.

The analysis can easily be extended to find PDFs of the momentum flux in the system of Eqs. (7) - (8), with the action $S_{\lambda}$ 
\begin{eqnarray}
S_{\lambda} & = & -i \int d\zeta \left(C_1 \dot{F}\bar{F}_1 + i C_2 k_{\xi}^2 F \bar{F}_1  + C_3 G F \bar{F}_2 + i \nu C_4 F \bar{F}_1 \right) \nonumber \\
& + & \frac{\kappa_0}{2} \int d\zeta \left( \bar{F}_1^2 + \bar{F}_2^2 \right) \nonumber \\
& + & i \lambda R_0 \int d\zeta F(\zeta)^2 \delta(\zeta) \nonumber \\
& - & i \int d\zeta \bar{F}_3 (D_2 G - R D_3 F^2 ),
\end{eqnarray}
where $R_0$ is the Reynolds stress
\begin{eqnarray}
R_0 & = & - \int d \xi \left( \psi_1 \frac{\partial \psi_1}{\partial \xi} \right).
\end{eqnarray}
Note that an average over $x$ has already been taken and that the momentum flux is non-zero since there is a natural phase shift introduced by the electron physics. A similar equation for the instanton can be found, but with different values of $F$ and $G$ at $t=0$,
\begin{eqnarray}
F_0 & = & \sqrt{\frac{2 i \kappa_0 \Phi_1}{C_1^2}} \sqrt{\lambda}, \\
G_0 & = & \frac{R D_3}{D_2} \frac{2 i \kappa_0 \Phi_1}{C_1^2} \lambda. 
\end{eqnarray}
The PDF tail of momentum flux can then be computed by following a similar analysis with the result
\begin{eqnarray}
P(R) & \sim & \exp\{-\Theta_M (\frac{R}{R_0})^{2}\}, \\
\Theta_M & = &  \frac{1}{2h_M},
\end{eqnarray}
where
\begin{eqnarray}
h_M = \frac{1}{4} \sqrt{\eta_2} + \frac{2 \kappa_0}{C_1^2}. 
\end{eqnarray}
Interestingly the exponential forms of the PDFs of momentum flux are qualitative different from what was found in ITG turbulence~\cite{a5}-~\cite{a7} and in Hasegawa-Mima turbulence~\cite{a3}-~\cite{a4} which was the stretched PDF $\sim \exp \{-c R^{3/2} \}$. 

Specifically the predicted PDF tails of momentum flux are Gaussian when zonal flow feeds back on the turbulence. This result differs from what has theoretically been found earlier~\cite{a7} where the zonal flow is treated kinematically. It is however in agreement with predictions from non-linear simulations of turbulence~\cite{a61}-~\cite{a62} and gyro-kinetic toroidal simulations~\cite{a91}-~\cite{a92}. The reason for this is the cubic non-linearity in Eq. (7) describing the ITG fluctuations while previously the zonal flow was treated as passively evolving by the turbulence. Physically, it is because the feedback of zonal flow regulates turbulence, limiting its growth from turbulence. It is interesting that although our model for turbulence and zonal flows is expected to have a limited region of validity due to the perturbation expansion, it captures the main features of what has been found in numerical simulation of plasma turbulence with zonal flows.

Note that non-linear models with cubic non-linearities for turbulent fluctuations was obtained using perturbation theory and are thus valid only for small values of the amplitude (small $\phi_1$). The statistical property of the fluctuations differs radically in systems with cubic and quadratic non-linearities as shall be discussed in more detail in Sec. VII. The predictive power of the method used here should be able to discriminate between models since the statistics of the fluctuations and zonal flow may change qualitatively. We note that, in ETG turbulence an evolution equation for the fluctuations with a cubic nonlinearity has previously been found~\cite{a93}. 

\section{The PDF tails in the reduced model two field model}
Due to the perturbation method, the five field model considered in Sec. III-IV is only valid for weak drift wave and zonal flow electric potential. Moreover an exact non-linear solution in this model is not available that can be used as an ansatz for a coherent structure for the intermittent transport. We will thus now consider the reduced two field model where a non-linear solution (modon) can be found and no reductive perturbation expansion is needed. It is a reduced model where the ITG mode turbulence is modeled using the continuity and temperature equation for the ions with adiabatic electrons~\cite{a6}-~\cite{a7}. In this model the effects of parallel ion motion, magnetic shear, trapped particles and finite beta on the ITG modes are neglected since they were shown to be not critical in previous work. Unlike the five-field model, the generation of a zonal flow is treated passively while the background fluctuations are affected by an imposed sheared velocity $V_0$. We note that, the effects of mean flow on the zonal flow are weak~\cite{a101}-~\cite{a102} while the effects of mean flow on the turbulence itself is much more prominent.

We formally calculate the PDF tails of momentum flux and zonal flow structure formation by using the instanton method. The probability distribution function for Reynolds stress $Z_1 = R$ or zonal flow structure formation $Z_2 = \phi_{ZF}$ can be defined as in a generalized Eq. (9) where 
\begin{eqnarray}
I_{\lambda_j} = \langle \exp(-i \lambda_j Z_j) \rangle.
\end{eqnarray}
The integrand can then be rewritten in the form of a path-integral as
\begin{eqnarray}
I_{\lambda_j} = \int \mathcal{D} \phi \mathcal{D} \bar{\phi} \mathcal{D} \phi_{ZF} \mathcal{D} \bar{\phi}_{ZF}  e^{-S_{\lambda_j}}.
\end{eqnarray}
Here, the parameter $j$ refers to the two specific cases included in the present study. The angular brackets denote the average over the statistics of the forcing $f$. By using the ansatz $T_i = \chi \phi$, the effective action $S_{\lambda_j}$ in Eq. (11) can be expressed as,
\begin{eqnarray}
S_{\lambda_j} & = & -i \int d^2x dt \bar{\phi} \left( \frac{\partial \phi}{\partial t} - (\frac{\partial }{\partial t} - \alpha_i \frac{\partial }{\partial y}) \nabla^2_{\perp} \phi + V_0(1-\nabla^2) \phi \right. \nonumber \\
& + & \left. (1-\epsilon_n g_i \beta)\frac{\partial \phi}{\partial y} - \beta [\phi, \nabla^2_{\perp} \phi ]\right) \nonumber \\
& + & \frac{1}{2} \int d^2x d^2 x^{\prime} \bar{\phi}(x) \kappa(x-x^{\prime}) \bar{\phi}(x^{\prime}) \nonumber \\
& + & i \lambda_1 \int d^2 x dt (-\frac{\partial \phi}{\partial x} \frac{\partial \phi}{\partial y}) \delta(t) \nonumber \\
& + & i \lambda_2 \int d^2 x dt \phi_{ZF} \delta(t) \nonumber \\
& + & \int d^2x dt (\dot{\phi}_{ZF} + R_0\langle v_x v_y \rangle).
\end{eqnarray}
In Eq. (51), 

\begin{eqnarray}
\beta & = & 1 + \frac{1}{\tau} + \frac{1}{\tau} \chi, \\
\chi & = & \frac{\eta_i - \frac{2}{3}(1-U + V_0)}{U - V_0 +\frac{5}{3\tau} \epsilon_n g_i}, \\
R_0 & = & \int d^2x \left( -\frac{\partial \psi}{\partial x} \frac{\partial \psi}{\partial y} \right).
\end{eqnarray}

The PDF tails are found by calculating the value of $S_{\lambda_j}$ at the saddle-point in the two cases; the PDF tail of momentum flux by taking into account the effect of a shear flow ($\lambda_1 \rightarrow \infty, \lambda_2 = 0$) and the PDF tail of structure formation of zonal flow ($\lambda_1 = 0, \lambda_2 \rightarrow \infty$). The integral in Eq (51) is divided in five parts; $K_1$ the ITG integral; $K_2$ the forcing integral; $K_3$ the momentum flux integral; $K_4$ the zonal flow integral; and finally $K_5$ represents the zonal flow evolution integral. The action in the first case can be found as;
\begin{eqnarray}
S_{\lambda_1} & \simeq & - \frac{1}{3} i h \lambda_1^3, \\
h & = & K_1 + K_2 + K_3 + K_4 + K_5, \\
K_1 & = & \frac{1}{2\kappa_0} \left( \gamma^2(1+\epsilon^2) [(\frac{4H_0}{H_0-1}-1)^{3/2}-1] \frac{C^{3/2}}{A} \right. \\
& +  &\left. 24 (1 + 6 \epsilon^2) \gamma_2^2 \frac{C^{3/2}}{A^{5/2}} (\frac{1}{3} \frac{H_0}{(H_0-1)^3} - \frac{1}{4} \frac{H_0}{(H_0-1)^2}) \right), \\
K_2 & = & \frac{1}{2\kappa_0} \left( \gamma^2(1+\epsilon^2) [(\frac{4H_0}{H_0-1}-1)^{3/2}-1] \frac{C^{3/2}}{2A} \right. \\
& +  &\left. 24 (\frac{1}{2} + 2 \epsilon^2) \gamma_2^2 \frac{C^{3/2}}{A^{5/2}} (\frac{1}{3} \frac{H_0}{(H_0-1)^3} - \frac{1}{4} \frac{H_0}{(H_0-1)^2}) \right), \\
K_3 & = & R_0 F^2(0), \mbox{       } K_4 = 0, \\ 
K_5 & = & \frac{2R_0 \sqrt{C}}{A} \frac{1}{H_0 - 1}, \\
H_0 & = & 4A-2, \mbox{       } A = \frac{\eta}{(1+\epsilon^2)\gamma^2}.
\end{eqnarray}
Here $C = 2 \kappa_0 \lambda_1^2/((1+\epsilon^2)\gamma)$, $\gamma = c_1 (1 + k^2 + 2 \alpha/k^3)$ ($k$ is the modon wave number), $c_1 = \alpha a /J_1(ka)$ ($a$ is the modon size), $\alpha = A_1 - k^2 A_2$, $A_1 = (1-\epsilon_n - U + V_0)/\beta$, $A_2 = (\alpha_i + U -V_0)/\beta$, $\gamma_2 = k \beta \alpha /2$, $\eta = (1+6 \epsilon^2) \gamma_2^2$ and $\kappa_0$ is the strength of the forcing $\kappa(x-y)$. To obtain the integrals in Eq. (51) we have used the localized modon solution, involving the parameter $\epsilon$ as done in Refs.~\cite{a6}-~\cite{a7}. The PDF tail of the Reynolds stress ($R$) can now be found by performing the integration over $\lambda_1$ in Eq. (51) using the saddle-point method in the same fashion as done in Ref.~\cite{a6}-~\cite{a7} i.e. recall Eq. (9) gives $P(R) \sim \int d \lambda_1 \exp\{-i \lambda_1 R - S_{\lambda_1}\} \sim \int d \lambda_1 \exp \{ - i \lambda_1 R + i h \lambda_1^3\}$. Now, the saddle point integral is evaluated at the saddle point $\lambda_{1 MAX} = \sqrt{R/(3 h)}$ which maximizes $P(R)$ with the result
\begin{eqnarray}
P(R) & \sim & \exp\{-\xi_1 (\frac{R}{R_0})^{3/2}\}, \\
\xi_1 & = &  \frac{2}{3}\frac{1}{\sqrt{3h}}.
\end{eqnarray}
This result is similar to the previous ones where a similar exponential PDF was found~\cite{a3}-~\cite{a7}. However, here the focus will be on how the coefficient $\xi_1$ depend on the imposed shear flow ($V_0$).

In the second limit where the limit of $\lambda_1 = 0 \ \mbox{and} \ \lambda_2 \rightarrow \infty$ gives the PDF tail of the structure formation itself, the action can be shown to be
\begin{eqnarray}
S_{\lambda_2} & \simeq & - i h \lambda_2^{3/2},
\end{eqnarray}
with
\begin{eqnarray}
K_3 & = & 0, \\
K_4 & = & 2 \phi_{ZF 0} \frac{\sqrt{C}}{A}, \\
C & = & 2 \kappa_0 \lambda_2/((1+\epsilon^2)\gamma).
\end{eqnarray}
The PDF tail can now be computed by following similar analysis, with the result
\begin{eqnarray}
P(\phi_{ZF}) & \sim & \exp\{-\xi_2 (\phi_{ZF}/\phi_{ZF0})^3\}, \\
\xi_2 & = & \frac{4}{27}\frac{1}{h^2} .
\end{eqnarray}

Recall that in the computation of the PDF tails of the momentum flux or zonal flow structure formation, the latter was assumed to be driven by a modon in the ITG turbulence. Since this modon is created by the forcing and that $F(t)=0$ as $t \rightarrow - \infty$, we could treat Eqs (66) and (70) as the transition amplitudes from an initial state with no fluid motion to a state with different values of momentum flux $R/R_0$ or zonal flow formation $\phi_{ZF}/\phi_{ZF 0}$. 

Note that all physical quantities are included in the parameters $\xi_1$ and $\xi_2$, through the ion temperature gradient ($\eta_i$), density gradient ($\epsilon_n$), temperature ratio ($\tau=T_i/T_e$), modon size ($a$), modon speed ($U$) and wave number ($k$). It is also important to note that $\xi_j \rightarrow \infty$ (i.e. PDF vanishes) as the the forcing disappears ($\kappa_0 \rightarrow 0$); the instanton cannot form and the PDF vanishes ($P(R) \rightarrow 0$ and $P(\phi_{ZF}) \rightarrow 0$) without forcing. 

\section{Numerical results of the reduced two field model}
We have presented a theory of the PDF tail of structure formation and how the PDF tail of momentum flux is modified by the presence of an imposed  shear flow ($V_0$). The exponential forms of the two PDF tails are completely different, signifying the difference in the physical interpretation. In the case of structure formation the PDF tails are found as $\sim \exp\{ -\xi_2 \phi_{ZF}^3\}$, while the momentum flux PDF tail $\sim \exp\{ -\xi_1 R^{3/2}\}$. The origin of these scalings are the quadratic non-linearity in the dynamical system (Eqs 1-2). Mathematically, the difference in scaling of the PDF tails momentum flux and zonal formation comes from the difference in the scaling of the initial condition ($F_0$) with the large parameter $\lambda$. The spatial structure of the modon is of less importance in determining the exponent in the exponential PDF tails than the temporal behavior, but changes the overall amplitude through the coefficients $\xi_1$ and $\xi_2$. Therefore an approximate spatial structure is sufficient to determine the exponential scaling whereas the correlation time of $f$ may affect the scaling. The spatial structure of the flow is incorporated in the initial condition of the flow $\phi_{ZF 0}$. In this section the parametric dependencies of $\xi_j$ will be studied in detail.

\begin{figure}
  \includegraphics[height=.3\textheight]{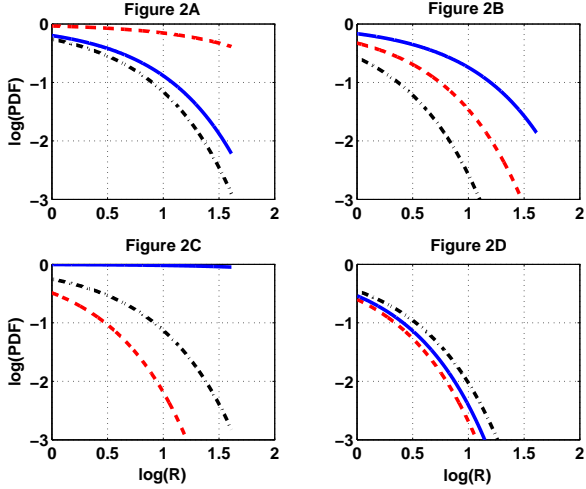}
  \caption{The PDF tail of momentum flux in ITG turbulence incorporating the effects of shear flow for $V_0 =0.0 $ (Figure 2A), $V_0 = 5.0$ (Figure 2B), $V_0 = 10.0$ (Figure 2C) and $V_0 = 15.0$ (Figure 2D) for $\eta_i = 2.0 $ (red, dashed line), $\eta_i = 4.0 $ (blue, solid line) and $\eta_i = 6.0 $ (black, dashed-dotted line).}
\end{figure}

First the PDF tail of momentum flux in ITG turbulence incorporating the effects of shear flow is shown in Figure 2. The parameters are $\tau = 2.0$, $\epsilon_n = 1.0$, $g_i = 1$, $a = 2$, $U= 2.0$, $\kappa_0 = 3000$, $\epsilon = 0.1$, $k \approx 1.91$ with $V_0 =0.0 $ (Figure 2A), $V_0 = 5.0$ (Figure 2B), $V_0 = 10.0$ (Figure 2C) and $V_0 = 15.0$ (Figure 2D) for $\eta_i = 2.0 $ (red, dashed line), $\eta_i = 4.0 $ (blue, solid line) and $\eta_i = 6.0 $ (black, dashed-dotted line). When $V_0 = 0$, the result recovers the previous finding in Ref.~\cite{a6}. It is clearly shown that the PDF tail of momentum flux is significantly reduced if a strong shear flow is present whereas weak flow can increase the PDF tail. In the equations the flow speed and the modon velocity come in as a combination of the form ($U-V_0$), determining the behavior of the resulting PDF tails. When ($U-V_0$) decreases, the PDF tail increases until it eventually decreases. This means that there exists a negative value ($U-V_0$) that gives maximum PDF tail, depending on all other parameters. Note that although the Galilean invariance of Eq. (1)-(2) may allow us to perform Galilean transformations to find other instanton solutions, those solutions correspond to different ground states in the field theory and must be discarded since they have nonzero velocity in the infinite past~\cite{a10}. Thus, our instanton solutions are not Galilean invariant~\cite{a10}.

\begin{figure}
  \includegraphics[height=.3\textheight]{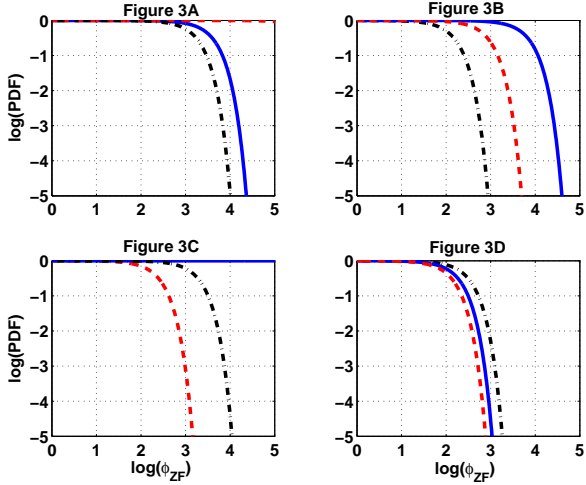}
  \caption{The PDF tail of structure formation in ITG turbulence incorporating the effects of shear flow for $V_0 =0.0 $ (Figure 2A), $V_0 = 5.0$ (Figure 2B), $V_0 = 10.0$ (Figure 2C) and $V_0 = 15.0$ (Figure 2D) for $\eta_i = 2.0 $ (red, dashed line), $\eta_i = 4.0 $ (blue, solid line) and $\eta_i = 6.0 $ (black, dashed-dotted line).}
\end{figure}
 
Second, the PDF tails of structure formation in ITG turbulence incorporating the effects of shear flow is shown in Figure 3. The parameters and the interpretation of the results are the same as those in Figure 2.

\section{Discussion and conclusions}
In summary, this paper presents the first prediction of the PDF tails of zonal flow formation from a dynamic system of turbulence and zonal flows. The PDF tails of zonal flow is $PDF \sim \exp \{ - c_{ZF j} \phi_{ZF}^3\}$ (where $j = 1,2$ represent the different constants depending on the models) which corroborate earlier findings~\cite{a3}-~\cite{a7}. 

In the first part a general five field model ($n_i, \phi, T_i, T_e, v_{i \parallel}$) where a reductive perturbation method is used to derive dynamical equations for drift waves and a zonal flow is studied. Interestingly the predicted PDF tails of momentum flux in the five field model of ITG turbulence are Gaussian when the feedback of zonal flow on the turbulence is incorporated. This result differs from what has theoretically been found earlier in Ref.~\cite{a7} and found in the reduced two field model of ITG turbulence in Sec. V whereas it is in agreement with predictions from non-linear simulations of turbulence~\cite{a61}-~\cite{a62} and in gyro-kinetic toroidal simulations~\cite{a91}-~\cite{a92}. Mathematically it is due to the different highest non-linearity between the two models (i.e. cubic in the five-field model and quadratic in the two field model). Physically, it is because in the two-field model the feedback of zonal flow on the fluctuations was not treated self-consistently, which will regulate turbulence, inhibiting its growth. 

Note that non-linear models with cubic non-linearities for the turbulence fluctuations was obtained perturbatively and thus valid only for small values of the amplitude (small $\phi_1$). The statistical property of the fluctuations differs in systems with cubic and quadratic non-linearities.  To elucidate these differences it is instructive to discuss a general formula for PDF tails of any moment in non-linear systems derived previously~\cite{a41}. By using the instanton method, the PDF of the m-th moment was shown to be
\begin{eqnarray}
P(Z) &  \sim & \exp\{ - c Z ^{s}\}, \\
s & = & \frac{n+1}{m} 
\end{eqnarray}
in the case where $n$ is the highest non-linearity in a dynamical system.

For instance, in the case of the five-field model, $n=3$ [Eq. (9)] while in the case of the two-field model, $n=2$ [Eq. (1)]. For PDFs of momentum flux, $m=2$ while for PDFs of zonal flow formation $m=1$. According to Eqs. (72)-(73) there is a significant difference in the PDF tails of the first moment in a system with cubic non-linearity where $s = 4$ and one with a quadratic non-linearity where $s=3$~\cite{a42}. For weak zonal flow Eq. (7) becomes linear and we expect Gaussian statistics ($s=2$) for the fluctuations. Note that for the five-field model, Eq. (72)-(73) suggest that the PDF tails of zonal flow formation are determined by the quadratic non-linearity in the dynamical equation for the zonal flow whereas the cubic non-linearity determines the PDF tails of momentum flux.

Finally, in the second part a reduced two-field model ($\phi, T_i$) that has an exact non-linear solution (bipolar vortex soliton) of ITG turbulence and zonal flows are studied to compute the PDFs of zonal flow formation and momentum flux. One of the important results from the numerical study in this reduced two-field model, which is also supported in the five field model, is that shear flows can significantly reduce the PDF tails of momentum flux and zonal flow formation. Since zonal flows are more likely to be generated further from marginal stability, they will then regulate ITG turbulence, leading to a self-regulating system. Namely, while ITG turbulence is a state with high level of heat flux, it also generates stronger zonal flows that inhibit transport. 

\section{Acknowledgment}
This research was supported by the Engineering and Physical Sciences Research Council (EPSRC) EP/D064317/1.

\appendix
\renewcommand\theequation{\thesection.\arabic{equation}}
\setcounter{equation}{0}
\section{Coefficients}
We have balanced the non-linear transport due to the imaginary parts of $\alpha$ [$\Im(\alpha)$] (A.15) with the source term $S_{Ti}$,
\begin{eqnarray}
S_{Ti} & + & 2 i k_y \left( \Im(\alpha) \frac{\partial}{\partial x}(\phi_1^{(5/3)*}\phi_1^{(1)} - \phi_1^{(5/3)}\phi_1^{(1)*}) \right. \nonumber \\
& - & \left. \Im(\frac{\partial \alpha}{\partial k_y}) \frac{\partial}{\partial x} (\phi_1^{(1)} \frac{\partial \phi_1^{(1)*}}{\partial \xi} - \phi_1^{(1)*} \frac{\partial \phi_1^{(1)}}{\partial \xi}) \right) = 0.
\end{eqnarray}
\begin{eqnarray}
C_1 & = & -k_y^3 (1+\tau) \epsilon_n ( 3 \tau^2 \epsilon_n [ 10 \epsilon_n + 3 \tau (2-3\eta_i)] \nonumber \\
& \times & \omega(-k_y \lambda + \omega)^2 + k_{\parallel}^2(25 k_y^3 \epsilon_n^2[5(1+\tau)\epsilon_n \nonumber \\
& + & \tau (-2 + 3 \eta_i)] \lambda^2 + 45 k_y^2 \tau \epsilon_n [ 5(1+\tau) \epsilon_n \nonumber \\
& + & \tau (-2 + 3 \eta_i)] \lambda^2 \omega + 27 k_y \tau^2 [ 5(1+\tau)\epsilon_n \nonumber \\
& + & \tau (-2 + 3 \eta_i)] \lambda^2 \omega^2 + \tau \{ 50 \epsilon_n^2 + 15 \tau \epsilon_n [ (2 - 3 \eta_i) + 4 \lambda] \nonumber \\
& + & 9 \tau^2 \lambda [ (4-6\eta_i) + 5 + 3 \tau)\lambda]\} \omega^3 ),\\
C_2 & = & k_y^3 L(1+\tau) \epsilon_n \omega (5 k_y \epsilon_n + 3 \tau \omega)(\tau \omega^2 \{ k_y^2 \epsilon_n [ 5(-7 \nonumber \\
& + & 5 D_{\perp}^2) \epsilon_n + 3 \tau (-2 + 3 \eta_i)] + 30 k_y (-1 \nonumber \\
& + & D_{\perp}^2) \tau \epsilon_n \omega + 9 (-1 + D_{\perp}^2 \tau^2 \omega^2\} \nonumber \\
& - &k_{\parallel}^2 [ 5 k_y^2 \epsilon_n[5(1+\tau)\epsilon_n + \tau (-2 + 3 \eta_i)] \nonumber \\
& + & 6 k_y \tau [ 5(1+\tau) \epsilon_n + \tau (-2 + 3 \eta_i)]\omega \nonumber \\
& + & 3 \tau^2 (5+3\tau)\omega^2]),\\
C_3 & = & \sqrt{\frac{2}{L_x}}(C_{NL1} k_m - 4 k_m^3 C_{NL2} \nonumber \\
& + & C_{NL3}[(k_m^2 + l^2 k_y^2)k_m - 4 k_m^3]),\\
C_{NL1} & = & \frac{1}{\lambda(5 \epsilon_n + 3 \tau \lambda)} \{ k_y^4 l^2 \tau \epsilon_n \omega^2 (5 k_y \epsilon_n + 3 \tau \omega) \nonumber \\
& \times & [ k_{\parallel}( k_y \{ -50 \epsilon_n^2 \lambda - 3 \tau^2 [ 10 \epsilon_n + 3 (2 - 3 \eta_i)]\lambda^2 \nonumber \\
& + & 9 \tau^3 (-2 + 3 \eta_i)\lambda^2 - 10 \tau \epsilon_n[3 \lambda^2 + 5 \epsilon_n (\eta_e + \lambda)] \nonumber \\
& - & 3 \tau^2 [ 10 \epsilon_n \eta_e + 3 (1+\tau)(-2+3\eta_i)\lambda]\omega) \nonumber \\
& + & \omega (k_y^2 \epsilon_n [175 \epsilon_n \lambda + 9 \tau^3 (2-3\eta_i)\lambda^2 \nonumber \\
& + & 35 \tau \epsilon_n [ 3 \lambda^2 + 5 \epsilon_n (\eta_e + \lambda)]+ 3 \tau^2 \lambda [3 (2-3 \eta_i)\lambda \nonumber \\
& + & 5 \epsilon_n (5 \eta_e + 7 \lambda)]\} + 3 k_y \tau \epsilon_n [ 50 \epsilon_n \lambda \nonumber \\
& + & 3 \tau^2 \lambda [ (-2 + 10 \eta_e + \eta_i) + 10\lambda ] \nonumber \\
& + & \tau [ 10 \epsilon_n (6 \eta_e + 5 \lambda) + 3 \lambda (-2 + 3 \eta_i + 10 \lambda)]\} \omega \nonumber \\
& + & 9 \tau^2(5 \epsilon_n + \tau \lambda)[ \lambda + \tau (\eta_e + \lambda)]\omega^2)]\},\\
C_{NL2} & = & \frac{1}{(5 \epsilon_n + \tau \lambda)}( k_y^4 l^2 \tau (1 + \eta_i + \tau \lambda)\omega^2( 5 k_y \epsilon_n \nonumber \\
& + & \tau \omega)^2 \{ 10 k_{\parallel}^2 \epsilon_n - \omega [ 5 k_y \epsilon_n(7 \epsilon_n + 3 \tau \lambda) \nonumber \\
& + & 3 \tau (5 \epsilon_n + 3 \tau \lambda)\omega]\}),\\
C_{NL3} & = & D_{\perp}^2 k_y^4 l^2 \tau(1 + \tau) \epsilon_n \omega^3 (5 k_y \epsilon_n + 3 \tau \omega)^3,\\
C_{4} & = & D_{\perp}^4 k_y^3 l \tau(1 + \tau) \epsilon_n \omega^3 (5 k_y \epsilon_n + 3 \tau \omega)^3,\\
D_1 & = & \frac{1}{(1+\tau) \epsilon_n \lambda^2 (5 \epsilon_n + 3 \tau \lambda)} \times ( 25 \epsilon_n^4 \eta_e \nonumber \\
& + & 5 \tau \epsilon_n^3 \lambda ( 6 \eta_e - 5 D_{\perp}^2 \lambda) + 9 D_{\perp}^2 \tau^2 \lambda^4 ( 1 \nonumber \\
& + & 2 \eta_i \tau \lambda) + D_{\perp}^2 \tau \epsilon_n \lambda^3 [ 10 (1+\eta_i) + ( 22 \nonumber \\
& - & 3 \tau)\tau \lambda] + \epsilon_n \lambda^2 \{3 \tau [2 + \tau(2 + 5 \eta_e - 3 \eta_i)-3 \eta_i] \nonumber \\
& + & 5 D_{\perp}^2 [7 (1+\eta_i) + 2(7 - 3 \tau)\tau \lambda]\}),\\ 
D_2 & = & -(3D_{\perp}^2 \tau \lambda^3(1 + \eta_i + \tau \lambda) + 5 \epsilon_n^3 [ \eta_e - (1+\tau)\lambda] \nonumber \\
& + & \epsilon_n^2 \lambda \{ [7 + \tau(7 + 10 \eta_e - 3 \eta_i) - 3 \eta_i + 5 D_{\perp}^2(1+\eta_i)] \nonumber \\
& + & \tau[5 D_{\perp}^2 - 3(1+\tau)]\lambda\} \nonumber \\
& + & \epsilon_n \lambda^2 [3 \tau (1+\tau + \tau \eta_e) + D_{\perp}^2(-7+3\tau)(1 + \eta_i + \tau \lambda)]) \nonumber \\
& \times & \frac{1}{\lambda(1+\tau)\epsilon_n},\\
D_3 & = & \sqrt{\frac{2}{L_x}}k_m l^2 ( k_y^2 D_{NL1} + D_{NL2}),\\
D_{NL1} & = & - \frac{\tau}{(1+\tau)\epsilon_n}[5 \epsilon_n^2 + (-7 + 3\tau)\epsilon_n \tau - 3 \tau \lambda^2],\\
D_{NL2} & = & 3 \epsilon_n k_y \Re \{ \frac{\partial \alpha }{\partial k_y}\},\\
\frac{ \partial \alpha}{\partial k_y} & = & (\frac{\omega}{k_y} - \lambda) \frac{\frac{10 \epsilon_n}{3 \tau}- (\eta_i - \frac{2}{3})}{k_y(\frac{5 \epsilon_n}{3 \tau} + \frac{\omega}{k_y})^2}, \\
\alpha & = &  \frac{\eta_i - \frac{2}{3} + \frac{2 \omega}{3 k_y}}{\frac{5 \epsilon_n}{3 \tau} + \frac{\omega}{k_y}},\\
\lambda & = & \frac{\omega}{k_y} - \frac{2}{k_y} \frac{\lambda_n}{\lambda_d}, \\
\lambda_n & = & \omega k_y^2 \{ \omega(1 + \frac{5}{3 \tau}) + k_y \Gamma\}, \\
\lambda_d & = & (1 + k_y^2(1 + \frac{5}{3 \tau}))\omega - \frac{k_y}{2}(1 - \epsilon_n [1 + \frac{5}{3 \tau} + \alpha_r]) \nonumber \\
& - & k_y^2 \Gamma - i [\frac{\epsilon_n s}{2 q} (1 + \frac{5}{3 \tau})], \\
\alpha_r & = & \frac{5}{3 \tau}, \\
\Gamma & = & \frac{1}{\tau}(\eta_i - \frac{2}{3}) + \frac{5}{3 \tau}\epsilon_n (1 + \frac{1}{\tau}).
\end{eqnarray}

\end{document}